\documentclass[fleqn]{2023SCGE}
\usepackage{hyperref}
\usepackage{orcidlink}
\usepackage{graphicx} 
\newcommand{\orcid}[1]{\href{https://orcid.org/#1}{\includegraphics[width=8pt]{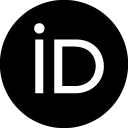}}}

\begin{document}

\ensubject{subject}

\ArticleType{Article}%??Article
\SpecialTopic{SPECIAL TOPIC: }%???????
\Year{2023}
\Month{January}
\Vol{66}
\No{1}
\DOI{??}
\ArtNo{000000}
\ReceiveDate{January 11, 2023}
\AcceptDate{April 6, 2023}
\newcommand{\roma}[1]{\uppercase\expandafter{\romannumeral#1}}
\newcommand{\speed}[1]{#1 km~s${}^{-1}$}
\newcommand{\density}[1]{#1 g~cm${}^{-3}$}
\newcommand{\accel}[1]{#1 km~s${}^{-2}$}
\newcommand{\acc}[1]{#1 m~s${}^{-2}$}
\newcommand{\nfig}[1]{Figure~\ref{#1}}
\newcommand{\rsun}[1]{${#1}\,R_\odot$}
\newcommand{\tbl}[1]{Table~\ref{#1}}
\newcommand{\mfig}[1]{Fig.\ref{#1}}
\newcommand{\mn}{{Mon. Not. R. Astron. Soc.}}
\newcommand{\mnras}{\mn}
\newcommand{\aj}{{Astron. J.}}
\newcommand{\apj}{{Astrophys. J.}}
\newcommand{\apjl}{{Astrophys. J. Lett.}}
\newcommand{\apjs}{{Astrophys. J. Supp.}}
\newcommand{\apss}{{Astrophys. and Space Sc.}}
\newcommand{\aaps}{{Astron. Astrophys. Supp.}}
\newcommand{\aapr}{{Astron. Astrophys. Rev.}}
\newcommand{\aap}{{Astron. Astrophys.}}
\newcommand{\nat}{{Nat.}}
\newcommand{\pasj}{{Publ. Astron. Soc. Jpn }}
\newcommand{\prd}{{Phys. Rev. D}}
\newcommand{\prl}{{Phys. Rev. Lett.}}
\newcommand{\pasp}{{Pub. Ast. Soc. Pac.}}
\newcommand{\procspie}{Proc. SPIE}
\newcommand{\ssr}{Space Science Reviews}
\newcommand{\solphys}{{\it Solar Phys.}}
\newcommand{\grl}{{\it Geophys. Res. Lett}}
\newcommand{\araa}{{ARA\&A}}
\newcommand{\na}{{New Astro.}}

\newcommand{\xp}[1]{{\textcolor{black}{#1}}}
\newcommand{\xpz}[1]{{{\textcolor{black}{#1}}}}
%\OnlineDate{January 1, 2016}
%%%%%%%%%%%%%%%%%%%%%%%%%%%%%%%%%%%%%%%%%%%%%%%%%%%%%%%

%%% title: ????
%%%   \title{title}{title for citation}
\title{Consecutive Narrow and Broad Quasi-periodic Fast-propagating Wave Trains Associated with a Flare}{ }

\author[1]{Xinping Zhou\orcid{0000-0001-9374-4380}}{{xpzhou@sicnu.edu.cn}}%
\author[2]{Yuandeng Shen\orcid{0000-0001-9493-4418}}{ydshen@ynao.ac.cn}
\author[2]{Chengrui Zhou\orcid{0009-0005-5300-769X}}{}%\protect\\?§Þ??§Ø????
\author[2]{Zehao Tang\orcid{0000-0003-0880-9616}}{}%
\author[3]{Ahmed Ahmed Ibrahim}{}

\AuthorMark{Xinping Zhou}

\AuthorCitation{X. P, Zhou, et al}

\address[1]{College of Physics and Electronic Engineering, Sichuan Normal University, Chengdu, 610068, People's Republic of China}
\address[2]{Yunnan Observatories, Chinese Academy of Sciences, Kunming 650216, People's Republic of China}
\address[3]{Department of Physics and Astronomy, College of Science, King Saud University, 11451 Riyadh, Saudi Arabia}

\abstract{The excitation mechanism of coronal quasi-period fast-propagating (QFP) wave trains remains unresolved. Using Atmospheric Imaging Assembly onboard the {\em Solar Dynamics Observatory} observations, we study a narrow and a broad QFP wave trains excited one after another during the successive eruptions of filaments \xpz{hosted} within a fan-spine magnetic system on 2013 October 20. The consecutive occurrence of these two types of QFP wave trains in the same event provides an excellent opportunity to explore their excitation mechanisms and compare their physical parameters. Our observational results reveal that narrow and broad QFP wave trains exhibit distinct speeds, periods, energy fluxes, and relative intensity amplitudes, although originating from the same active region and being associated with the same {\em GOES} C2.9 flare. \xp{Using} wavelet analysis, we find that the narrow QFP wave train shares a similar period with the flare itself, suggesting its possible excitation through the pulsed energy release in the magnetic reconnection process that generated the accompanying flare. On the other hand, the broad QFP wave train appears to be associated with the energy pulses released by the successive expansion and unwinding of filament threads. Additionally, it is plausible that the broad QFP wave train was also excited by the sequential stretching of closed magnetic field lines driven by the erupting filament. These findings shed light on the different excitation mechanisms and origins of the QFP wave trains.}



\maketitle

\begin{multicols}{2}
\section{Introduction}\label{section1}
Magnetohydrodynamic (MHD) waves are ubiquitous in the magnetized solar atmosphere and have garnered significant attention from solar physicists in recent decades \xpz{\cite{2005LRSP....2....3N,2020SSRv..216..136L,2020ARA&A..58..441N,2022SCPMA..6539611L}}, primarily due to their importance in understanding puzzling phenomena, such as the solar atmosphere heating and the acceleration of the fast solar wind \cite{1999Sci...285..862N,2020ARA&A..58..441N, 2020SSRv..216..136L, 2020SSRv..216..140V,2021ApJ...908..233S}. Furthermore, MHD waves play a crucial role in diagnosing the physical properties of the solar atmosphere using seismology techniques \cite{1970PASJ...22..341U, 1984ApJ...279..857R, 2005SSRv..121..115N, 2022SoPh..297...20S}. 

Large-scale extreme-ultraviolet (EUV) waves manifest as a single wavefront traveling at typical speeds ranging from \speed{200} to \speed{1500} \cite{2013ApJ...776...58N}. These waves can propagate across a large fraction of the solar disk \cite{2014SoPh..289.3233L, 2015LRSP...12....3W, 2020shen_ChSB}. In certain cases, a pair of wavelike features can be observed, with the preceding wavefront's speed being approximately three times that of the following one \cite{2011ApJ...732L..20C, 2012ApJ...752L..23S, 2014ApJ...786..151S, 2021SoPh..296..169Z, 2022ApJ...939L..18S}. Previous studies show that the preceding component is a fast-mode magnetosonic wave or a shock wave \cite{2020MNRAS.493.4816M}, exhibiting wave effects such as transmission, reflection, and refraction during its interaction with coronal magnetic structures like active regions \cite{2012ApJ...752L..23S, 2013ApJ...773L..33S}, coronal holes \cite{2012ApJ...746...13L, 2013ApJ...775...39Y}, as well as the coronal magnetic null-point \cite{2023arXiv231017573Y}. Furthermore, large-scale EUV waves can often result in the oscillations of remote filaments and coronal loops \cite{2012ApJ...754....7S, 2014ApJ...786..151S, 2014ApJ...795..130S, 2017ApJ...851..101S}. However, the physical nature of the following component is still an open question. Some authors have explained the slower component as a pseudo wavelike phenomenon caused by the reconfiguration of coronal magnetic fields \cite{2002ApJ...572L..99C,2005ApJ...622.1202C}, while others suggest it could be a slow-mode wave \cite{2009ApJ...700.1716W, 2012SCPMA..55.1316M}. Notably, the magnetic field line stretching model proposed by Chen et al. \cite{2002ApJ...572L..99C} can adequately explain the simultaneous existence of the fast and slow wavelike components in an EUV wave event. The authors proposed that the faster component is a fast-mode magnetosonic wave or a shock driven by a coronal mass ejection (CME), while the slower apparent wavelike front is formed due to the successive stretching of the closed confining magnetic field lines of an erupting magnetic flux rope \cite{2002ApJ...572L..99C}.

Recently, high\xpz{-resolution} \xp{spatial and temporal} EUV observations taken by the Atmospheric Imaging (AIA) \cite{2012SoPh..275...17L} instrument onboard the {\em Solar Dynamics Observatory} (SDO) \cite{2012SoPh..275....3P} have revealed the existence of quasi-periodic fast-propagating (QFP) wave trains. These wave trains consist of multiple concentric and coherent arc-shaped wavefronts propagating at a fast-mode magnetosonic speed ranging from several hundreds to over \speed{2000} \cite{2011ApJ...736L..13L,2012ApJ...753...53S,2014SoPh..289.3233L}. The discovery of coronal QFP wave trains provides a new seismological tool for diagnosing the coronal property. However, the generation mechanism and propagation properties of QFP wave trains are not yet fully understood, necessitating further observational and theoretical studies. According to different physical properties revealed by a statistical study, Shen et al. \cite{2022SoPh..297...20S} proposed that the QFP wave trains could be classified into two categories, i.e., narrow and broad QFP wave trains. Following is a brief introduction to the two types of QFP wave trains.

\subsection{Narrow QFP Wave Trains}
The narrow QFP wave train was \xp{first} reported by Liu et al. \cite{2011ApJ...736L..13L} using high spatial and temporal resolution observations taken by the {\em SDO}/AIA. These wave trains propagates at a high speed along open or closed coronal loops. According to a recent statistical study conducted by Shen et al. \cite{2022SoPh..297...20S}, the \xp{phase} speed, deceleration, angular width, amplitude, and the period of narrow QFP wave trains are in the ranges of \speed{305-2349}, \accel{0.1-5.8}, 10$^\circ$-80$^\circ$, 1\%-8\% and 25-550 s, respectively. The energy flux carried by narrow QFP wave trains is estimated to be on the order of $\approx 10^5$ erg cm$^{-2}$ s$^{-1}$, which is sufficient to heat the local coronal loops \cite{2011ApJ...736L..13L, 2018ApJ...860...54O, 2022SoPh..297...20S}. Typically, narrow QFP wave trains are observed in AIA 171 \AA\ channel, although some cases can be detected simultaneously in the AIA 193 and 211 \AA\ channels \cite{2010ApJ...723L..53L, 2013SoPh..288..585S}. The preference for the AIA 171 \AA\ channel can be attributed to factors such as the temperature of the wave-hosting plasma, the small intensity amplitude of narrow wave trains, and the high photon response efficiency of the AIA 171 \AA\ channel \cite{2016AIPC.1720d0010L}. 

Narrow QFP wave trains differ significant from large-scale single-pulsed EUV waves \cite{2011JASTP..73.1096Z, 2013ApJ...776...58N,2013A&A...554A.144Y, 2018ApJ...864L..24L, 2019ApJ...870...15L, 2023ApJ...949L...8Z}. They exhibit a close physical relationship with the accompanying flares, often sharing similar periods and having a close temporal and spatial association \cite{2012ApJ...753...53S, 2013SoPh..288..585S, 2018ApJ...853....1S}. In some cases, narrow QFP wave trains can repeatedly occur along the same loop system, with each wave train accompanied by an energy burst in the flare \cite{2013A&A...554A.144Y, 2020ApJ...889..139M, 2022ApJ...941...59Z}, suggesting a strong connection between the generation of narrow QFP wave trains and energy releasing process in flares. They can also co-occur along different (or bi-directional) coronal loops rooted in the same eruption source region \cite{2021ApJ...908L..37M}, or simultaneously along open and closed loops \cite{2011ApJ...736L..13L}. \xp{Zhang et al. first reported a QFP wave simultaneously detected with slow wave apparently propagating along a funnel coronal loop system, interpreted as co-existing fast and slow magnetoacoustic waves excited by different mechanisms \cite{2015A&A...581A..78Z}.} 

Several explanations have been proposed for the generation of QFP wave trains \cite{2022SoPh..297...20S}. However, the most likely physical mechanisms are the dispersive evolution \cite{1983Natur.305..688R, 2013A&A...560A..97P, 2014A&A...569A..12N, 2017ApJ...847L..21P, 2018MNRAS.477L...6S} and the pulsed energy releasing in magnetic reconnection processes \cite{2011ApJ...736L..13L, 2012ApJ...753...53S, 2013A&A...554A.144Y, 2018ApJ...868L..33L, 2018ApJ...855...53L, 2022ApJ...941...59Z}. The dispersion evolution mechanism refers to the dispersive evolution of an impulsively generated broadband disturbance, which naturally leads to the generation of a QFP wave train in \xp{a waveguide} at a distance from the initial site. This occurs because a propagating fast-mode magnetosonic wave with different frequencies travels at different group speeds \cite{1983Natur.305..688R,1984ApJ...279..857R}. Therefore, in this scenario, the period of a wave train is primarily determined by the physical parameters of the waveguide and its surroundings \cite{1983Natur.305..688R, 2004MNRAS.349..705N}. \xp{ In the scenario of \xpz{the} pulsed energy releasing mechanism, the period of the wave train is determined by the disturbance source. Observations have revealed that the oscillation period in a narrow wave train is similar to that of the quasi-periodic pulsations (QPPs) patterns in the associated flare \cite{2011ApJ...736L..13L, 2012ApJ...753...53S, 2013SoPh..288..585S, 2018ApJ...853....1S, 2013A&A...554A.144Y}. In principle, the flare can trigger a wave train via the production of QPPs by the magnetic reconnection. Until now, the generation mechanism \xpz{for} the QPPs in the associated flare is still an open issue. They could be interpreted \xpz{in terms of} quasi-period magnetic reconnection \cite{2006SoPh..238..313C, 2015ApJ...807...72L};   on the other hand, some QPPs in the flaring energy releases could \xpz{be} triggered by the leakage of 3-min chromospheric umbral oscillation \cite{2009A&A...505..791S}. However, it is worth noting that not all periods of the observed wave \xpz{trains} are consist with the range of the QPPs, which has been identified in recent studies \cite{2018MNRAS.477L...6S,2018MNRAS.480L..63S, 2018ApJ...860L...8S}. Both cases when the QFP wave period is independent of the QPP in a flare and coincides with it are \xpz{compatible} with the dispersion mechanism. In the former case, the dispersion leads to a quasi-periodic wave train from a broadband initial pulse, while in the latter case the wave driver is narrowband, which is not subject to the dispersive evolution.}

\subsection{Broad QFP Wave Trains}
In contrast to narrow QFP wave trains that travel along coronal loops, broad QFP wave trains usually propagate along the solar surface where the magnetic field has a strong vertical component. In other words, the propagation of narrow and broad QFP wave trains are parallel and perpendicular to the magnetic field, respectively. Broad QFP wave trains are typically observed by in all EUV channels of AIA and generally result in a large intensity amplitude of about 30\% relative to the background corona; they continuously emanate from the flare kernel with a large angular extent of about 90$^\circ$-360$^\circ$ and propagate outward at a speed of \speed{370-1100} and with a period in the range of 36-240\,s \cite{2014SoPh..289.3233L, 2022SoPh..297...20S}. Furthermore, broad QFP wave trains are capable of propagating over significant distance, similar to large-scale EUV waves, and often interact with various coronal structures such as filaments, coronal loops, and coronal holes \cite{2012ApJ...753...52L, 2019ApJ...873...22S, 2022A&A...659A.164Z, 2022ApJ...930L...5Z}. During these interaction, wave effects such as reflection, transmission, and refraction can be observed clearly, providing strong evidence for the wave nature of broad QFP wave trains. Notably, Zhou et al. \cite{2022A&A...659A.164Z} firstly reported the occurrence of total reflection of a broad QFP wave train during its interaction with a remote polar coronal hole, which differs from previous observations where the incoming wavefronts were able to transmit through the coronal hole \cite{2022ApJ...930L...5Z}. Detailed analysis indicates that the total reflection of a QFP wave train requires the incidence and critical angles to satisfy the theory of total reflection, meaning that the incidence angle must be \xpz{greater} than the critical angle. Additionally, homologous broad QFP wave trains originating from the same active region but different eruptions have been observed, providing valuable insights into the condition favorable for the occurrence of QFP wave trains \cite{2022ApJ...936L..12W}.

The physical characteristics and evolution behaviors of broad QFP wave trains differ significantly from those of narrow QFP wave trains. However, they share similarities with large-scale single-pulsed EUV waves, except for the number of wavefronts. Large-scale single-pulsed EUV waves are generally believed to be driven by coronal mass ejections (CME) \cite{2006ApJ...641L.153C, 2011ApJ...738..160M, 2012SoPh..281..187P}. Nevertheless, recent observations have shown that many EUV waves are not necessarily associated with CMEs. For example, they can be driven by small-scale loop expansions caused by mini-filament eruptions \cite{2017ApJ...851..101S, 2023A&A...674A.167L}, or directly by coronal jets \cite{2013ApJ...764...70Z, 2018ApJ...861..105S,2022ApJ...929L...4Z}, and indirectly by sudden loop expansions due to the impingement of coronal jets \cite{2013MNRAS.431.1359Z,2018ApJ...860L...8S, 2018MNRAS.480L..63S, 2021RSPSA.47700217S, 2022ApJ...931..162Z}. Explaining the periodicity of broad QFP wave trains using the generation mechanisms of large-scale EUV waves is challenging, even though the two types of waves share some similar properties. Studies by Liu et al. \cite{2012ApJ...753...52L} and Zhou et al. \cite{2022ApJ...930L...5Z} have found that broad wave trains often exhibit a dominant period that matches the pulsation period of the accompanying flares. Therefore, they proposed that broad QFP wave trains are likely excited by the periodic energy release processes in flares, such as the periodic coalescence and splitting of plasmoids. On the other hand, Shen et al. \cite{2019ApJ...873...22S} analyzed a broad QFP wave train and found its period to be significantly different from that of the accompanying flare, making it difficult to explain using pulsed energy-releasing processes in flares. Instead, they discovered that the period of the wave train was similar to that of untwisting threads in an erupting filament within the eruption source region. Consequently, they proposed that the observed QFP wave train was possibly excited by pulsed energy release resulting from the periodic expansion of the initially twisted filament threads. Direct evidence for pulsed energy release in flares was reported by Shen et al. \cite{2022A&A...665A..51S}, who observed a broad QFP wave train in white-light observations in the outer corona, ranging from 2 to 4 solar radii. They observed the periodic generation and rapid expansion of strongly bent newly formed reconnection loops in a failed breakout eruption. Recently, Sun et al. \cite{2022ApJ...939L..18S} found in their observations that a broad QFP wave train might be excited by the successive stretching of magnetic field lines during a solar eruption. Since the speed of \xpz{inner edge of the QFP wave train} is approximately one-third of the QFP wave train speed, the authors argued that their observation is consistent with the magnetic field line stretching model proposed by Chen et al. \cite{2002ApJ...572L..99C}.

In recent years, numerous simulation studies have been conducted alongside observations to investigate the generation of QFP wave trains. For instance, Pascoe et al. \cite{2013A&A...560A..97P, 2014A&A...568A..20P} and Li et al. \cite{2018ApJ...855...53L} studied the evolution of impulsively generated narrow QFP wave trains in a funnel geometry loop and coronal holes. The effects of the transverse plasma density structuring on the formation and evolution of narrow QFP wave trains has also been explored in various works \cite{2018ApJ...855...53L}. Ofman et al. \cite{2011ApJ...740L..33O} generated a narrow QFP wave train in a magnetic funnel by applying periodic velocity pulsations at the low coronal boundary. The authors found that their simulated QFP wave trains exhibited physical parameters comparable to those observed in real observations. Regarding the excitation of broad QFP wave trains, several studies have considered the nonlinear physical process in magnetic reconnection to generate large-scale QFP wave trains. In this manner, broad QFP wave trains can be spontaneously produced without any artificial exciters \cite{2015ApJ...800..111Y, 2016ApJ...823..150T, 2021ApJ...911L...8W}. Additionally, some simulation studies have demonstrated that large-scale broad QFP wave trains can also be produced through the nonlinear steepening of the leaky wave train from the waveguide \cite{2017ApJ...847L..21P, 2023ApJ...943L..19S}. This scenario finds support in the sole observation reported by Nistic\'o et al. \cite{2014A&A...569A..12N}, and implies that narrow and broad QFP wave trains could be generated by a same physical process. The occurrence of either narrow or broad QFP wave trains in real observation could be an observational effect. If a narrow (broad) QFP wave train is detected, the corresponding broad (narrow) wave train should also be present.

As one of the significant new discoveries made by the {\em SDO}/AIA, coronal QFP wave trains have expanded the repertoire of coronal waves and present an excellent opportunity to deepen our understanding of coronal \xp{MHD} waves. However, the mechanisms responsible for exciting narrow and broad QFP wave trains have yet to be fully explored. Therefore, it is crucial to study more events to identify their excitation mechanisms and understand what determines the periods of these wave trains. In this paper, we present observations of a narrow and a broad QFP wave train that were both associated with a single flare event. The simultaneous occurrence of these two types of QFP wave trains offers an excellent opportunity to compare their physical parameters and investigate their generation mechanisms. The following section presents the analysis results, while the discussions and conclusions are provided in the final section.

\section{Observations and Results}
\label{se:results}
On 2013 October 20, a {\em GOES} X-ray C2.9 flare occurred in NOAA active region 11868, accompanied by two successive filament eruptions and a coronal mass ejection (CME). The flare started at 08:33 UT and reached its \xp{peak} at 08:40 UT. The filament eruptions in this event, along with their physical connection, was discussed in detail in previous study by Zhou et al. \cite{2021ApJ...923...45Z}. They emphasized that the two successive filament eruptions were physically connected to each other, and the physical linkage between the filament eruptions could be the magnetic implosion mechanism \cite{2000ApJ...531L..75H, 2012ApJ...750...12S}. In the present work, we revisit the event to explore the excitation mechanisms of the consecutive appearance of a narrow and a broad QFP wave trains that are not analyzed in Zhou et al. \cite{2021ApJ...923...45Z}. Additionally, we compare the physical parameters and evolution characteristics of the two types of QFP wave trains. \xp{The pixel size and cadence of the AIA images are 0.6$^{\prime\prime}$ and 12\,s,} respectively \cite{2012SoPh..275...17L}, which are calibrated using the standard program \texttt{aia\_prep.pro} in the solar software and are differentially rotated to a reference time 08:40:00 UT. The soft X-ray 0.5-4.0 \AA, 1.0-8.0 \AA\ and 173.2 MHz fluxes provided by {\em GOES} and Nan$\c{c}$ay Radioheliograph \cite{kerdraon2007nanccay} were used to analyze the fine structure of the flare pulsations, respectively. The Kanzelh{\"o}he Solar Observatory (KSO) provides the full-disk at 6562.8 \AA\ with a temporal cadence of 1 minute and a spatial resolution of 2$^{\prime\prime}$. We note that in the present event, the narrow \xp{and broad QFP wave trains can be detected in the AIA 171 \AA\ channel (dominated by emission from Fe {IX}, sensitive to the 0.8 MK plasma), but is weak and almost invisible in other bands.}

\begin{figure}[H]
	\centering
	\includegraphics[scale=0.28]{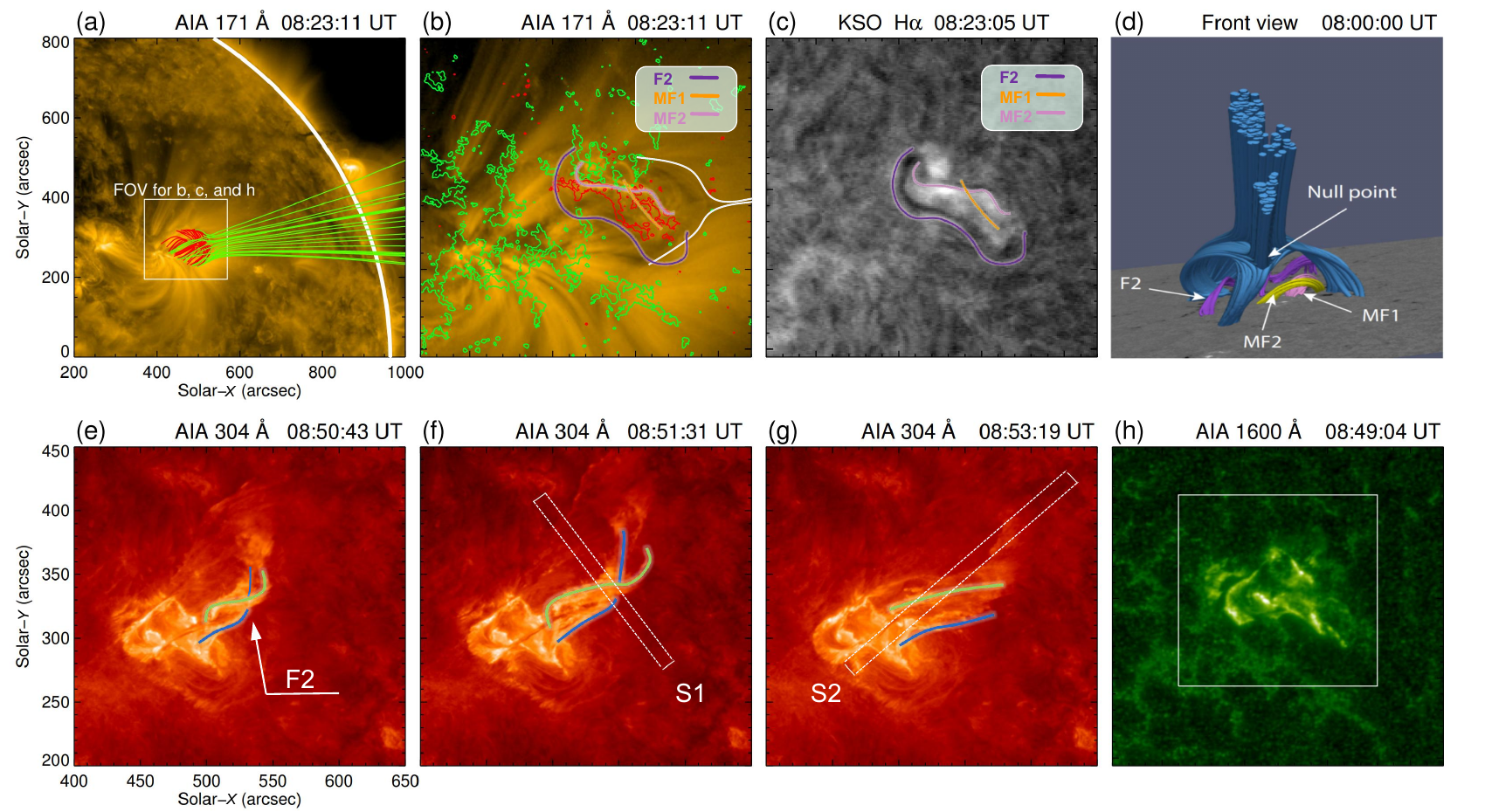}
	\caption{Panel (a) shows the initial coronal condition of the eruption source region using the {\em SDO}/AIA 171 \AA\ direct images, in which the red and green lines represent the closed and open magnetic field lines of the fan-spine structure extrapolated by the PFSS, and the white box indicates the field of the view of panels (b) and (c). The red and green curves in panel (b) indicate the positive and negative polarities at $\pm 50$ G, while the white curves outline the approximate location of the spine. The purple, yellow, and pink curves represent the filaments F2, MF1, and MF2 profiles, respectively. Their locations are also overlaid on the AIA 171 \AA\ image in panel (b). Panel (d), taken from Figure 7 of Zhou et al. \cite{2021ApJ...923...45Z} and obtained the original author's permission, displays the overlying fan-spine structure and the filaments underneath. Panels (e), (f) and (g) show the eruption process of the filament F2, where the rectangles labeled S1 and S2 are used to obtain the time-distance stack plots shown in \cref{fig:tdp}, \xp{and the green and blue curves depict the profiles of the untwisting filament F2. } Panel (h) exhibits the flare ribbons observed in 1600 \AA\, where the white box is used to collect AIA light curves in \cref{fig:wavelet}.} 
	\label{fig:overview}
\end{figure}

\cref{fig:overview} (a) presents the pre-eruption environment of the source region using the AIA 171 \AA\ direct image taken at 08:23:11 UT. In this image, selected closed and open magnetic field lines from the Potential-Field Source-Surface \xp{(PFSS)} \cite{1969SoPh....6..442S} model are overlaid as red and green lines, respectively. \xp{PFSS model is designed to visualize the solar coronal magnetic field and has been widely utilized in solar physics \cite{2011NewA...16..276B,2013ApJ...773..162B,2015ApJ...805...48B,2023ApJ...949...66L,2023ApJ...944..116L,2023MNRAS.518..388K,2021ApJS..257...34F}}. It is evident that the source region forms a fan-spine magnetic system, consisting of inner closed (red) and outer fan-spine field lines (green). Panel (b) is the close-up view of the white box in panel (a). The red and green contours represent the positive and negative magnetic polarities, respectively. The eruption source region was composed of a positive magnetic area surrounded by negative magnetic fields, a typical configuration for forming a fan-spine magnetic system in the low corona \cite{2019ApJ...885L..11S}. The KSO H$\alpha$ center image in panel (c) displays two crossing mini filaments, MF1 and MF2, and a large filament, F2. Their profiles are marked with yellow, pink, and purple curves, respectively. Combined analysis of the filament positions in the KSO H$\alpha$ image and the magnetic field information shown in panel (b), it is evident that the filaments were located on the magnetic polarity reversion lines. Panel (d) is adapted from Figure 7 in Zhou et al. \cite{2021ApJ...923...45Z}, illustrating the magnetic field lines extracted by the nonlinear force-free field (NLFFF) at 8:00 UT. This panel cofirms that all the filaments were situated under the fan structure, consistent with the observation in panels (a) and (b). Additionally, the inclination angle $\beta$ of the outer spine relative to the solar surface is estimated to be approximately 60$^\circ$ (also refer to Figure 7 (e) in Zhou et al. \cite{2021ApJ...923...45Z} for a different view). MF1 and MF2 initially merged into a long filament, F1, through magnetic reconnection around the crossing site. Subsequently, at about 08:30 UT, F1 erupted. Approximately 15 minutes later, F2 started to rise and underwent a violent eruption triggered by the disturbance resulting from the eruption of F1 . During the eruption, the main body of F2 exhibited noticeable untwisting motion (see blue and green curves in panels (e)-(g) and the Supplementary Materials). \xp{This evolution is consistent with the explanation proposed by Canfield et al. \cite{1996ApJ...464.1016C}: As the untwisting motion of the filament, the filament will appear both blueshifts and redshifts and the tight filament gradually relaxes. (The detailed evolution of the erupting filament F2 refer to the online animation)}. The flare ribbons associated with the filament eruptions are shown in panel (h), observed using the AIA 1600 \AA\ channel.

\begin{figure}[H]
	\centering
	\includegraphics[scale=0.3]{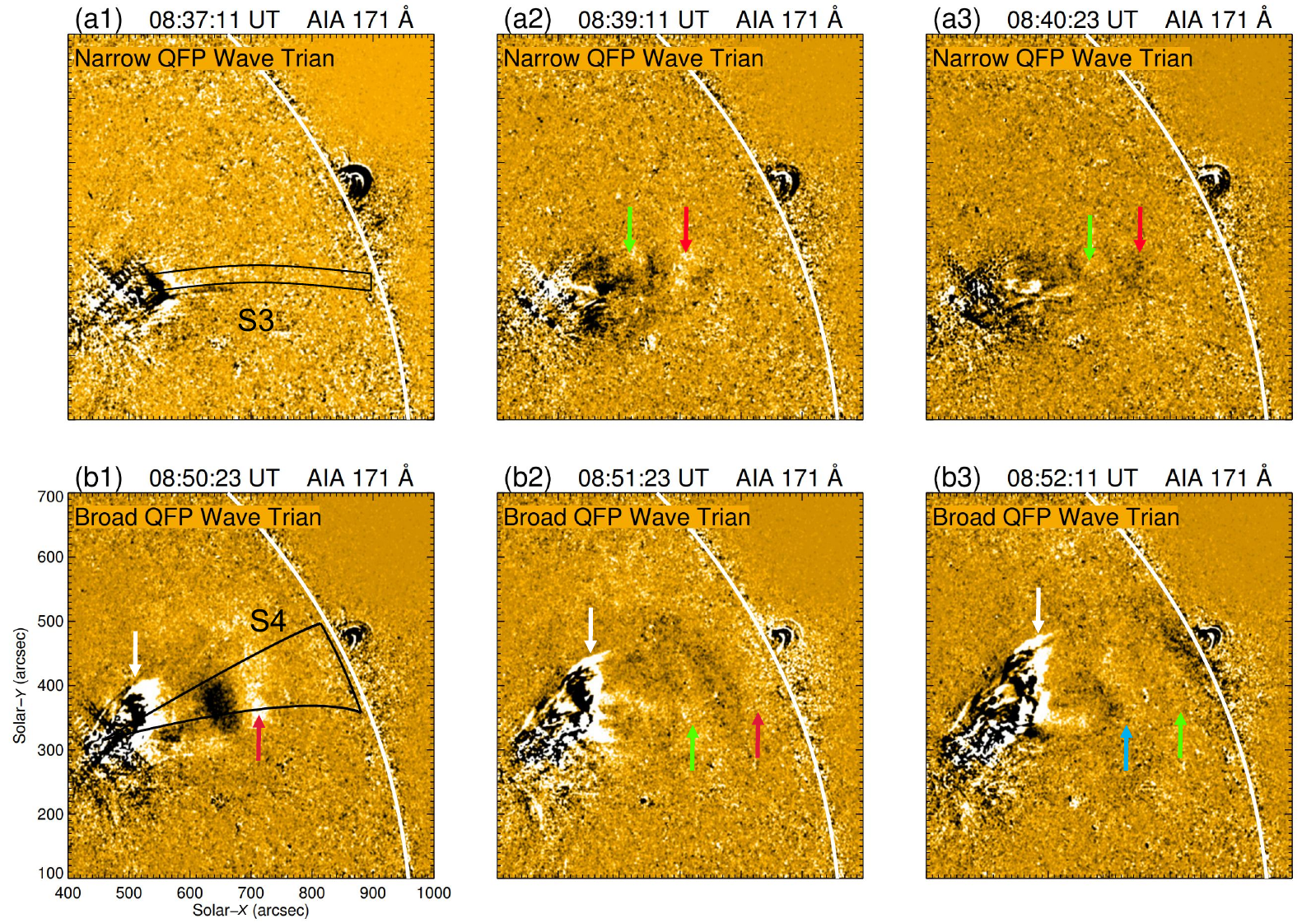}
	\caption{Panels (a1)-(a3) and (b1)-(b2) show the snapshots of the narrow and broad QFP wave trains at different times using the AIA 171 \AA\ running difference images. \xp{The red, green and blue arrows marked the positions of the first, second and third wavefront } of the narrow and broad QFP wave trains, respectively, \xp{while the white arrow point to the erupting filament}. The rectangle marked S3 in panel (a1) and the \xp{sector (angle 13$^\circ$)} denoted by S4 are used to reconstruct the time-distance stack plots in \ref{fig:tdp} to trace the evolution of the narrow and broad QFP wave trains. (An animation of this figure is in the Supplementary Materials.)} 
	\label{fig:evolution}
\end{figure}

During the eruption of F1, a series of wavefronts (red and green arrows in \ref{fig:evolution} (a2) and (a3)) were observed to emanate successively from the flaring region at 08:35 UT. These wavefronts rapidly propagated outward along the open loops, which constitute the outer spine of the fan-spine system (refer to \cref{fig:evolution} (a1) and (a2)). The wavefronts comprised a narrow wave train with an angular width of approximately 25$^\circ$. After a few minutes, when the narrow QFP wave train disappeared from the AIA 171 \AA\ images, another set of wavefronts emerged during the eruption of F2, around 08:48 UT (red, green, and blue arrows in \cref{fig:evolution} (b1)-(b3)). \xp{These wavefronts originate at the periphery of the flare region, propagate in front of the erupting filament F2, and are immediately followed by coronal dimming, a typical feature of EUV waves \cite{2016GMS...216..381C,chen2022}}. As illustrated in \cref{fig:evolution} (b1)-(b3), these wavefronts propagated along the quiet Sun with an angular width of about 60$^\circ$, which is a significant difference path compared to that of the narrow QFP wave train that along the loops of the spine. The detailed evolution of the narrow and broad QFP wave trains can be found in the Supplementary Materials.

\begin{figure}[H]
	\centering
	\includegraphics[scale=0.325]{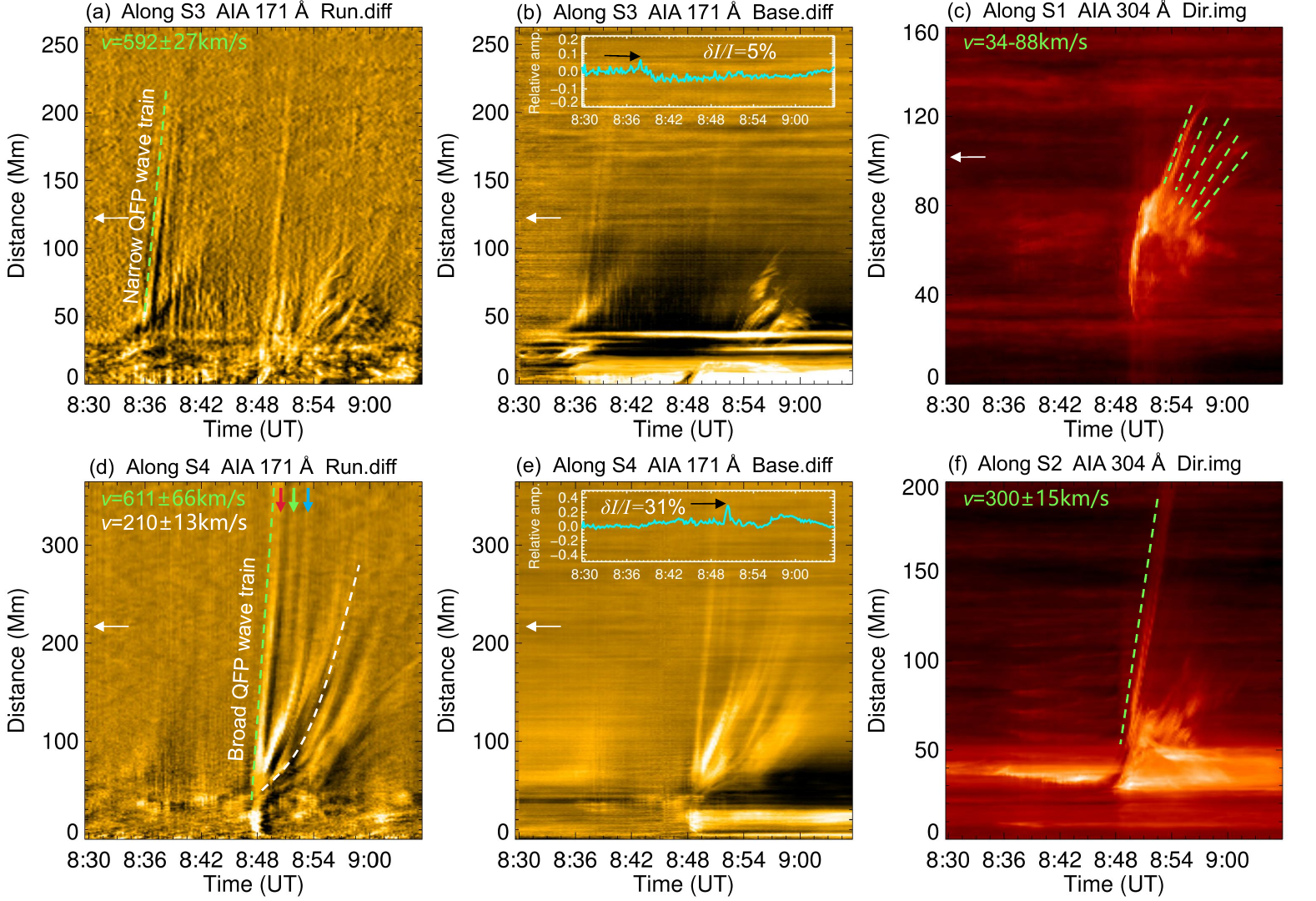}
	\caption{Panels (a) and (b) are time-distance stack plots respectively obtained from AIA 171 \AA\ running- and base-difference images along rectangle S3, while Panels (d) and (e) are along \xp{sector} S4. The insets in panels (b) and (e) are the time profiles of the relative amplitude extracted from the positions 120 Mm and 218 Mm in corresponding panels. Panels (c) and (f) display the untwisting and radial motion of the eruption filament F2 along rectangles S1 and S2, respectively. The green and white dotted lines depict the linear and second-order polynomial fittings for estimating the speeds of the moving feature, and the results are listed in each corresponding panel with different colors. The white arrows in panels (a)-(e) point to the positions where the intensity profile is used to analyze the periodicity and obtain the relative intensity amplitude, while the black arrows in panels (b) and (e) marked locations of the largest relative intensity amplitude of the narrow and broad QFP wave trains, respectively. \xp{The red, green and blue arrows in panel (d) indicate the ridges that correspond to the first, second and third wave fronts of broad QFP wave train, respectively.}  } 
	\label{fig:tdp}
\end{figure}

To study the kinematics of the wave trains, we made time-distance stack plots (TDSPs) using the AIA 171 \AA\ running-difference and base-difference images along the paths marked as S3 and S4 in \cref{fig:evolution} (a1) and (b1). The results are displayed in \cref{fig:tdp}. The bright tracks in the TDSPs represent the moving features, and their slopes indicate the speeds. Since the narrow and the broad QFP wave trains are respectively propagating along the coronal loops and the solar surface, the rectangular S3 is a projection on the plane of the sky along the narrow QFP wave train's trajectory, while the sector S4 with a \xp{13$^\circ$} wide angle is along the solar surface. As shown in the TDSPs, it can be found that most of the wavefronts of the narrow QFP wave lasted for less than 3 minutes and less than 5 minutes for the broad QPF wave. The onset times of the narrow and the broad QFP wave trains were about 08:35 UT and 08:48 UT, and their projection speeds were estimated to be around 592$\pm$\speed{27} and 611$\pm$\speed{66}, respectively. Considering the projection effect, the corrected speed $v_c$ of the narrow QFP wave train is about \speed{1184} ($v_c=v/\cos\beta,~\beta\approx 60^\circ$).

 In addition, the eruption of F2, labeled with a white dotted curve in panel (d), exhibited a significant acceleration process and reached a speed of 210$\pm$\speed{13}. Panels (b) and (e) show the TDSPs obtained from the AIA 171 \AA\ base-difference images along S3 and S4, respectively, providing information about the amplitude of the wave trains. Although the wavefront signals in panels (b) and (e) appear weaker compared to panels (a) and (d), the characteristics of the wavefronts can still be discerned. The upper-left corner in panels (b) and (e) displays the profile of the relative intensity amplitudes of the narrow and broad QFP wave trains, extracted at a distance of 120 Mm and 218 Mm from the origin of coordinates of panels (b) and (e), respectively. The result indicates that the largest relative intensity amplitude $\delta I/I$ of the narrow and the broad QFP wave trains were about 5\% and 31\%, respectively, consistent with the statistical results presented by Shen et al. \cite{2022SoPh..297...20S}. To analyze the eruption of the filament F2, the TDSPs created along the directions perpendicular and parallel to the eruption direction (marked with S1 and S2, respectively, in \cref{fig:overview}), as shown in \cref{fig:tdp} (c) and (f). In panel (c), the tracks marked with green dotted lines represent the unwinding filament threads. By fitting the tracks with a \xp{linear} function, we obtain that its transverse expansion speed was in the range of 34-\speed{88}. Similarly, we obtain its radial eruption speed is about \speed{300}. This speed is higher than that obtained along S4 because the eruption direction of F2 was along S2 rather than S4 in the plane of the sky.

\begin{figure}[H]
	\centering
	\includegraphics[scale=0.28]{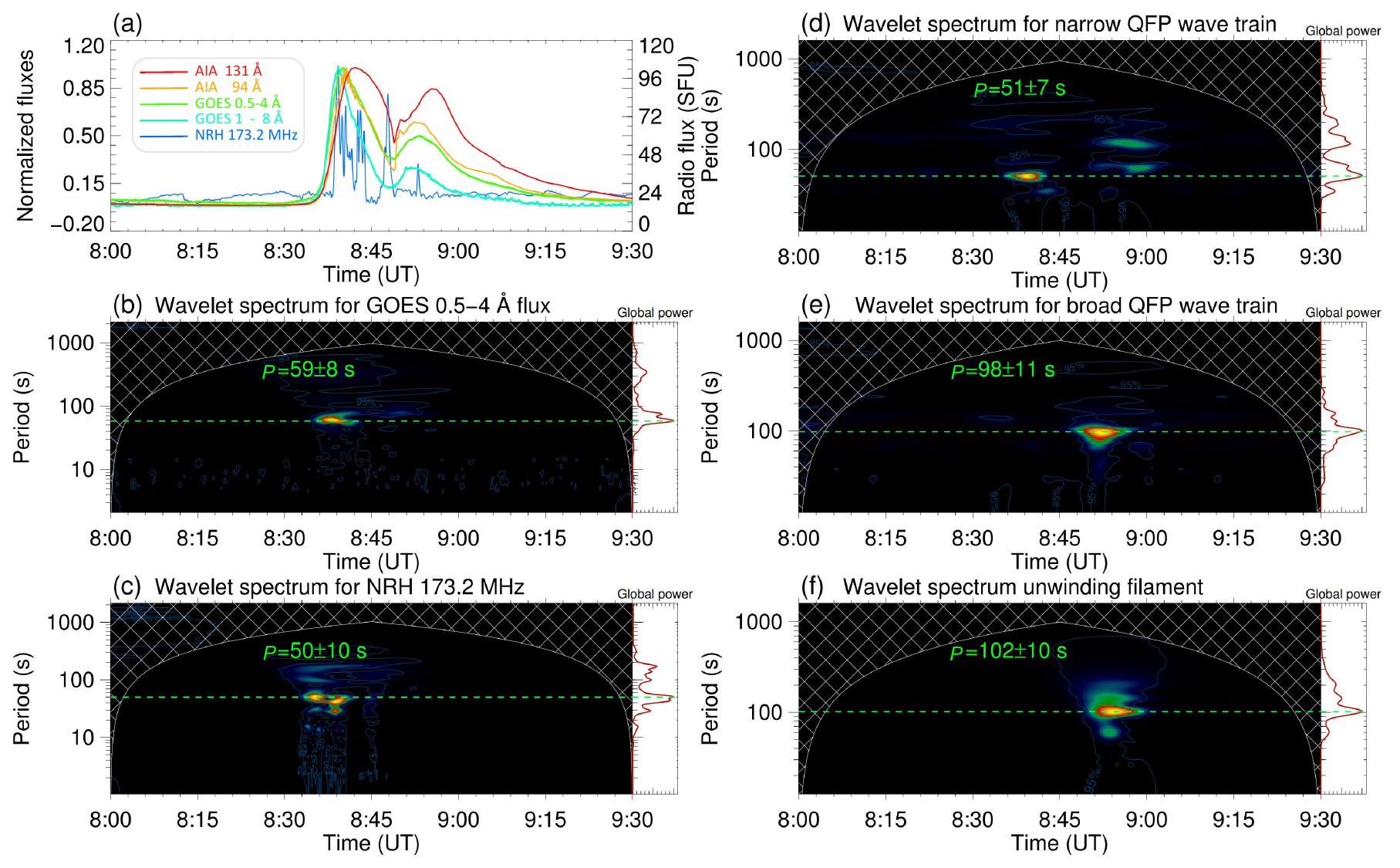}
	\caption{Panel (a) displays the normalized {\em GOES} 0.5-4.0 \AA\ and \AA\ 1-8 \AA\ X-ray flux curve, the NRH 173.2 MHz radio emission flux curve, and normalized AIA 94 \AA\ and 131 \AA\ light flux curve within the eruption source region outlined by the white box in \cref{fig:overview} (h). Panels (b) and (c) show the wavelet spectrum of the flare QPPs using the {\em GOES} 0.5-4.0 \AA\ and NRH 173.2 MHz detrend curves as an input, respectively. Panels (d)-(f) are, respectively, the wavelet spectrum for narrow QFP wave train, broad QFP wave train, and untwisting motion of the filament F2, using the extracted intensity profiles at the distances labeled with white arrows in \cref{fig:tdp} (a), (d) and (c), respectively. \xp{The specific method of extracting the intensity profile is to extract a rectangle region with a width of 10 pixels at the position marked by the white arrow in the TDSP, and then average the intensity in the $y$ direction (width direction) of the rectangle to obtain the distribution of intensity in time.} In each wavelet spectrum map, the period is highlighted by a green horizontal dashed line, and the corresponding period $P$ is also listed in the figure.} 
	\label{fig:wavelet}
\end{figure}

To explore the origin of these two wave trains, we employ the wavelet analysis method \cite{1998BAMS...79...61T} to derive the periods of the wave trains, the accompanying flare, and the unwinding filament threads of F2. For the flare, we use the soft X-ray and radio fluxes recorded by {\em GOES} and NRH to analyze the fine structure of the flare pulsations (see \cref{fig:wavelet} (a)). From the {\em GOES} fluxes, we find that the C2.9 flare was followed by a slight bump from about 08:48 UT to 08:57 UT, corresponding to the two successive filament eruptions. Comparing the AIA 131 \AA\ and 94 \AA\ light curves measured from the flaring region (see the white box in \cref{fig:overview} (h)), one can see that they have a similar trend with the {\em GOES} fluxes. Hence, we can utilize the {\em GOES} and NRH fluxes to explore the nonthermal energy release process associated with the accompanying flare \cite{2021ApJ...921L..33Y, 2022ApJ...941...59Z}. Using the detrended time derivative of the {\em GOES} soft X-ray flux in the energy band of 0.5-4.0 \AA\ as the input for the wavelet procedure, we obtain its wavelet spectrum, which is displayed in \cref{fig:wavelet} (b). One can identify that the main period 59$\pm$8 s appeared during the interval of the C2.9 flare. We obtained the same result by analyzing the {\em GOES} 1-8 \AA\ flux. To further validate the reliability of this result, we proceed to examine the periodicity of the flare using the NRH 173.2 MHz data and find a period of about 50$\pm$10s, similar to that obtained from the {\em GOES} fluxes. Notably, the timing of the flare pulsations derived from the NRH 173.2 MHz flux is consistent with that obtained from the {\em GOES} flux. These results indicate the analysis of the period of the flare is reliable.

For the narrow and the broad QFP wave trains, we first extracted the intensity profile at the distances of 120 Mm and 218 Mm from the origin of the coordinates in panels \cref{fig:tdp} (a) and (d), respectively. Subsequently, by using the detrended curves as inputs, the wavelet power spectrums reveal that the dominant periods of the narrow and the broad QFP wave trains were respectively 51$\pm$7\,s and 98 $\pm$11\,s (see \cref{fig:wavelet} (d) and (e)). It is clear that the period of the narrow QFP wave train is consistent with those revealed by the {\em GOES} and NRH radio fluxes\xp{. Thus, we speculate that generation of the narrow QFP wave train might have a relationship with the pulsed energy release in the accompanying flare}. In the wavelet spectrums, one can see that the duration time of the broad QFP wave train coincides with that of the small bump seen in the {\em GOES} fluxes after the main C2.9 flare. However, during this time interval, there is no periodic signal can be found in the wavelet spectrums generated from the {\em GOES} soft X-ray and NRH radio fluxes (see \cref{fig:wavelet} (b) and (c)). This suggests that the generation of the broad QFP wave train may not be associated with the pulsed energy release process in the accompanying flare. Next, we further analyze the periodicity of the unwinding threads of the erupting filament F2. Using the same method, we firstly extract the intensity profile at a distance of 100 Mm from the origin of the coordinates in \cref{fig:tdp} (c) and then calculate the wavelet spectrum as shown in \cref{fig:wavelet} (f). The result indicates that the period of the unwinding threads of the erupting filament F2 is 102$\pm$10 s, which is the same as the period of the broad QFP wave train. Besides, the duration period of the unwinding filament threads is also similar to that of the broad QFP wave train. The close temporal and period relationships between the unwinding threads of the erupting filament F2 and the broad QFP wave train imply that the latter might be excited by the successive expansion of the former, confirming the result proposed by Shen et al. \cite{2019ApJ...873...22S}.

The energy flux $\mathcal{F} $ carried by a wave can essentially be estimated by using $\mathcal{F}=\frac{1}{2}\rho v_{am}^2 v_{gr}\approx\frac{1}{2}\rho v_{am}^2 v_{ph}$, where $\rho$ is the plasma density, $v_{am}$ is the disturbance amplitude of the locally perturbed plasma,$v_{ph}$ and $v_{gr} $ are the phase speed and group speed of the wave train. Here we take the group speed $v_{gr}$ equal to the phase speed $v_{ph}$ as a rough estimate. Considering that the emission intensity is proportional to the square of the plasma density in the optically thin corona\xp{\footnote{\xp{The estimation used here is relatively crude, primarily due to the dependencies of the intensity on both column depth of the emitting plasma structure and the potential impact form the destructive interference of positive and negative perturbations of the density along the ray path \cite{2012A&A...543A..12G,2013A&A...555A..74A}.}}}, i.e., $I \propto \rho^2$, and combining the relationship $v_{am}/v_{ph} \geqslant \delta\rho/\rho=\delta I/(2I)$, the energy flux can be rewritten as $\mathcal{F}\geqslant\frac{1}{8}\rho v_{ph}^3(\frac{\delta I}{I})^2$ \cite{2011ApJ...736L..13L,2018ApJ...860...54O,2022ApJ...930L...5Z,2022ApJ...941...59Z}. For the present case, we use the phase speed (\speed{1184}) and the relative intensity amplitude (5\%) of the narrow QFP wave train and take the mean number density of coronal loops $n_e \approx 3.44\times 10^8$ cm$^{-2}$ s$^{-1}$ \cite{2021ApJ...908L..37M}, then we can derive the lower limit value of the energy flux of the narrow QFP wave train to be $\mathcal{F}\geqslant 3.4\times 10^5 $ erg cm$^{-2}$ s$^{-1}$. Similarly, we obtain the lower limit value of the energy flux of the broad QFP wave train as $\mathcal{F}\geqslant 7.5\times10^5 $ erg cm$^{-2}$ s$^{-1}$, where we take the number density of the quiet Sun corona as $n_e \approx 1.5\times 10^8$ cm$^{-2}$ s$^{-1}$ \cite{2004psci.book.....A}, and the phase speed and relative intensity amplitude of the broad QFP wave train as \speed{600} and $31\%$, respectively.

\section{Discussion and conclusions} 
\label{se:discussion}
We studied the consecutive generation of a narrow and a broad QFP wave train associated with two successive filament eruptions in a fan-spine magnetic system. The eruption event was accompanied by a {\em GOES} C2.9 flare, which exhibited a slight bump structure during its decay phase. Detailed analysis results revealed that the impulsive phase of the flare was primarily caused by the first eruption of the two mini filaments, while the second eruption of large filament F2 resulted in the small bump during the flare's decay phase. The flare exhibited strong pulsations in its impulsive phase, with periodic signal ranging approximately between 50 - 60 s. However, during the time interval of the small bump, no periodic signal was detected \xpz{in the light curves}. Wavelet analysis results indicated that the narrow QFP wave train has a similar period with the flare pulsation, indicating that the narrow QFP wave train exhibited a similar period to the flare pulsations, suggesting that it was possibly excited by the pulsed energy release process within the flare. On the other hand, the period of the broad QFP wave train correlated with the untwisting motion of the thin threads in the second erupting filament F2, indicating that the broad QFP wave train was likely excited by the untwisting filament threads. These two wave trains displayed distinct propagating preferences and temperature response ranges, and their physical parameters, such as speed, angular width, relative intensity amplitude, and energy flux, exhibited significant difference. These observational evidence strongly implies that these two wave trains are not the same type \cite{2022SoPh..297...20S}. Detailed parameters of the two QFP wave trains are listed in Table \ref{tab:parameters}.

\begin{table}[H]
	\footnotesize
	\begin{threeparttable}\caption{Physical parameters of the narrow and the broad QFP wave trains}\label{tab:parameters}
		\doublerulesep 0.1pt \tabcolsep 4pt 
		\begin{tabular}{ccc}
			\toprule
			{Parameters} & {Narrow QFP wave} & {Broad QFP wave} \\\hline
			Start time (UT) &  08:35 &  08:48 \\
			End time (UT)   &  08:39 &  08:55  \\
			Wavefront number  & 8 & 3  \\
			Wavelength (Mm)  & 69 & 97  \\ % & 69[41] & 97[101]  \\
			Period (s)  & 51$\pm$7 & 98$\pm$11  \\
			Speed (km s$^{-1}$)  & 1184 & 611  \\
			Angular width($^\circ$)  &25 & 60  \\
			Intensity Amplitude  & 5\% &31\%  \\
			Energy Flux ($\times10^{5}$ erg cm$^{-2}$s$^{-1}$)  & 3.4 & 7.5  \\
			\bottomrule
		\end{tabular}
	
\end{threeparttable}
\end{table}

Solar flares, which are among the most powerful energy release processes in the solar system, always exhibit highly variable emissions on timescales of sub-seconds to minutes in all wavelength bands from radio to X-ray. These variations, collectively referred to as QPPs, are commonly observed during the impulsive phase of flares \cite{2009SSRv..149..119N, 2018SSRv..214...45M, 2020STP.....6a...3K, 2020ApJ...888...53L, 2023SCPMA..6659611Z}. As a periodic phenomenon associated with flares, most of QFP wave trains have a close temporal and periodic relationship with QPPs in the flares \cite{2011ApJ...736L..13L, 2012ApJ...753...53S, 2013SoPh..288..585S, 2018ApJ...853....1S, 2022ApJ...941...59Z, 2022A&A...659A.164Z}, and the two periodic phenomena could be regarded as the two different aspects of the same physical process in flares \cite{2022SoPh..297...20S}. Therefore, studying the excitation mechanism of QFP wave trains can draw inspiration from the extensive research on those QPPs in flares \cite{2018ApJ...855...53L,2018ApJ...868L..33L,2020STP.....6a...3K,2021SSRv..217...66Z}. According to previous studies by Liu et al. \cite{2014SoPh..289.3233L} and Shen et al. \cite{2022SoPh..297...20S}, two dominant mechanisms are proposed for generation of QFP wave trains, i.e., the pulsed energy excitation mechanism and dispersion evolution mechanism. The pulsed energy excitation mechanism mainly associate with some nonlinear physical process in magnetic reconnection, such as the periodic generation, coalescence and ejections of plasmoids in current sheet \cite{2000A&A...360..715K, 2015ApJ...799...79N,2016ApJ...832..195N, 2016ApJ...823..150T,2018ApJ...853L..18Y,2022NatCo..13..640Y}, oscillatory reconnection \cite{2009A&A...493..227M, 2012ApJ...749...30M, 2017ApJ...844....2T, 2019ApJ...874..146H,2019ApJ...874L..27X,2019A&A...621A.106T, 2022ApJ...925..195K, 2022ApJ...933..142K, 2023ApJ...943..131K}, patchy magnetic reconnection \cite{2006ApJ...642.1177L, 2019MNRAS.489.3183C, 2020ApJ...905..165R}, oscillation of current sheet caused by super-Alfv\'{e}nic beams \cite{2016ApJ...829L..33L}, Kelvin-Helmholtz instability \cite{2006ApJ...644L.149O} and external disturbances from lateral or lower atmosphere layers \cite{2006A&A...452..343N, 2006SoPh..238..313C, 2012ApJ...753...53S, 2017ApJ...844..149K, 2018ApJ...868L..33L}. These nonlinear physical processes can individually or collectively generate QPPs and QFP wave trains. Recent simulations considering the pulsed energy release mechanism have  successfully produced narrow and broad QFP wave trains that exhibit similar parameters derived from real observations \cite{2011ApJ...740L..33O, 2015ApJ...800..111Y, 2016ApJ...823..150T}. 

The dispersion evolution mechanism, on the other hand, involves the dispersive evolution of a broadband disturbance that is impulsively generated along waveguides. Since waves with different frequencies \xpz{travel} at different speeds, an initial broadband wave packet can eventually evolve into a wave train along the waveguide. This mechanism has also been tested in a few simulations, which produces QFP wave trains as detected in observations \cite{2013A&A...560A..97P, 2014A&A...568A..20P, 2017ApJ...847L..21P}. Another characteristic of the dispersion evolution mechanism is that the period of the wave train should show a drift from high to low, resulting in the appearance of a tadpole-like structure in the time-dependent wavelet power spectrum \cite{2004MNRAS.349..705N, 2021MNRAS.505.3505K}.
Considering the QFP wave trains observed in the present study, the narrow one could be caused by nonlinear physical processes in the magnetic reconnection that results in the appearance accompanying C2.9 flare. This inference is based on the narrow QFP wave train exhibiting a consistent period of about 50 - 60 s with the flare over time. In addition, the wavelet spectrum of the narrow wave train does not display a tadpole-like feature, which is characteristic of the dispersion evolution mechanism. Therefore, it is unlikely that the narrow QFP wave train is generated through the dispersion evolution mechanism, and instead, the pulsed energy excitation mechanism involving nonlinear processes in magnetic reconnection is a more plausible explanation for its generation.

% In the present case, no narrow QFP wave train was simultaneously observed in any waveguide, and the flare did not exhibit any pulsation during the occurrence of the broad QFP wave train. Therefore, the generation of the broad QFP wave train in this case can not be explained by a combination of nonlinear physical processes in magnetic reconnection that produced the flare and the dispersion evolution of an impulsively generated broadband disturbance.
%

The excitation mechanism of the broad QFP wave is still not fully understand. Some authors have suggested that these waves may be excited by the nonlinear processes in flares \cite{2012ApJ...753...52L, 2021SoPh..296..169Z, 2022A&A...659A.164Z, 2022ApJ...930L...5Z, 2021ApJ...911L...8W, 2022A&A...665A..51S}, as they share the same period with the accompanying flares. However, other observations have showed that the periods of the broad QFP wave trains can differ significantly from those of the accompanying flares \cite{2019ApJ...873...22S}. \xp{In addition, numerical simulations conducted by Nistic\'o et al.\cite{2014A&A...569A..12N} and Pascoe et al. \cite{2017ApJ...847L..21P} have revealed that the dispersion evolution of an impulsively generated disturbance, \xpz{through the leakage of fast waves from the waveguide}, can also produce the broad QFP wave trains \cite{2023ApJ...943L..19S}. In particular, their simulation demonstrates that both narrow (guided) and broad (leaky) QFP wave trains can coexist during an eruption event. However, such a scenario is rarely observed in realistic observational detection due to its strong dependence on the observing angle. Consequently, expecting the simultaneous detection of guided (narrow) and leaky (broad) wave trains is highly unrealistic.}

Recently, researchers such as Shen et al. \cite{2019ApJ...873...22S} and Sun et al. \cite{2022ApJ...939L..18S} proposed that the generation of broad QFP wave trains does not necessarily require the appearance of QPPs in the accompanying flares. For example, Shen et al. \cite{2019ApJ...873...22S} found that the period of the broad QFP wave train in their case was similar to the unwinding period of the filament threads of the erupting filament. They concluded that their broad QFP wave train was likely excited by the periodic energy pulsed released from the sequentially unwinding and expanding filament threads. In our case, we also observed a common period about 100 s between the the broad QFP wave train and the erupting filament. Therefore, we proposed that the broad QFP wave train in our case should be excited by the untwisting threads of the erupting filament F2, supporting the findings of Shen et al. \cite{2019ApJ...873...22S}. Additionally, Sun et al. \cite{2022ApJ...939L..18S} found that the speed of the broad QFP wave train in their observation was about three times faster than the \xpz{apparent propagation of the inner edge of the QFP wave train}. They suggested that their broad QFP wave train was excited by the successive expanding of coronal loops, following the scenario proposed in the magnetic field line stretching model \cite{2002ApJ...572L..99C}. In our case, we also found a similar speed ratio between the broad QFP wave train and the expanding loops. However, we did not observe a similar period between the expanding loops and the broad QFP wave train. Therefore, further observational and theoretical investigations are required to test the scenario proposing the excitation of broad QFP wave trains by the successive stretching of closed magnetic field lines in a filament eruption, as suggested by Sun et al. \cite{2022ApJ...939L..18S}. \xpz{It would also be interesting to check whether there are chromospheric counterpart of the QFP wave train in the CHASE H$\alpha$ data \cite{2022SCPMA..6589602L,2022SCPMA..6589603Q}.}

%%%%%%%%%%%%%%%%%%%%%%%%%%%%%%%%%%%%%%%%%%%%%%%%%%%%%%%
%%% Acknowledgements. ??§Ý
%%%%%%%%%%%%%%%%%%%%%%%%%%%%%%%%%%%%%%%%%%%%%%%%%%%%%%%
\Acknowledgements{The authors thank the teams of {\em SDO}, {\em GOES} and NRH for providing the excellent data. This work was supported by the Natural Science Foundation for Youths of Sichuan Province (2023NSFSC1351), Natural Science Foundation of China (12303062, 12173083, 11922307), the Yunnan Science Foundation for Distinguished Young Scholars (202101AV070004), and the National Key R\&D Program of China (2019YFA0405000). We gratefully acknowledge ISSI-BJ for supporting the international team `Magnetohydrodynamic wave trains as a tool for probing the solar corona'. Co-author, Ahmed Ahmed Ibrahim would thank the Researchers Supporting Project number (RSPD2023R993), King Saud University, Riyadh, Saudi Arabia. }

%%%%%%%%%%%%%%%%%%%%%%%%%%%%%%%%%%%%%%%%%%%%%%%%%%%%%%%
%%% Conflict of interest. ????????????
%%%%%%%%%%%%%%%%%%%%%%%%%%%%%%%%%%%%%%%%%%%%%%%%%%%%%%%
\InterestConflict{The authors declare that they have no conflict of interest.}

%%%%%%%%%%%%%%%%%%%%%%%%%%%%%%%%%%%%%%%%%%%%%%%%%%%%%%%
%%% Supplements. ????????, ????
%%%%%%%%%%%%%%%%%%%%%%%%%%%%%%%%%%%%%%%%%%%%%%%%%%%%%%%
%\Supplements{}

%%%%%%%%%%%%%%%%%%%%%%%%%%%%%%%%%%%%%%%%%%%%%%%%%%%%%%%
%%% Reference section. ?¦Ï?????
%%% citation in the content using "some words~\cite{1,2}".
%%% ~ is needed to make the reference number is on the same line with the word before it.
%%%%%%%%%%%%%%%%%%%%%%%%%%%%%%%%%%%%%%%%%%%%%%%%%%%%%%%
\bibliographystyle{scpma}{}
%\bibliography{euv_wave}

%%%%%%%%%%%%%%%%%%%%%%%%%%%%%%%%%%%%%%%%%%%%%%%%%%%%%%%
%%% Appendix sections. ??????, ????
%%%%%%%%%%%%%%%%%%%%%%%%%%%%%%%%%%%%%%%%%%%%%%%%%%%%%%%

%\section{Name}

%\end{appendix}

%\begin{appendices}
%\section{Appendix}
%\end{appendices}
%\appendix

%\appendix

\end{multicols}
\end{document}